\documentclass[aip,reprint]{revtex4-1}

\usepackage{amsmath,amssymb}
\usepackage{graphicx}
\usepackage{hyperref}

\def\imply{\Rightarrow}

\newcommand{\dd}[2]{\frac {\partial #1}{\partial #2}}

\newcommand{\pdr}{\partial}
\newcommand{\DD}[2]{\frac {d #1}{d #2}}
\newcommand{\grad}{{\bf \nabla}}

\newcommand{\beq}{\begin{equation}}
\newcommand{\eeq}{\end{equation}}
\newcommand{\beqs}{\begin{eqnarray}}
\newcommand{\eeqs}{\end{eqnarray}}

\newcommand{\half}{\frac{1}{2}}
\newcommand{\ov}[1]{\frac{1}{#1}}

\def\del{\delta}	
\def\eps{\epsilon} 
\def\la{\lambda}		
\def\sig{\sigma}
\def\tht{\theta}
\def\om{\omega}		
\def\Om{\Omega}
\newcommand{\G}{{\Gamma}}

\newcommand{\bfS}{{\bf S}}

\newcommand{\bfP}{{\bf P}}

\newcommand{\bff}{{\bf f}}
\newcommand{\bfg}{{\bf g}}
\newcommand{\bfh}{{\bf h}}
\newcommand{\bfu}{{\bf u}}
\newcommand{\bfv}{{\bf v}}
\newcommand{\bfw}{{\bf w}}
\newcommand{\bfx}{{\bf x}}
\newcommand{\bfy}{{\bf y}}

\newcommand{\bfj}{{\bf j}}
\newcommand{\bfM}{{\bf M}}

\newcommand{\bfr}{{\bf r}}

\newcommand{\bfA}{{\bf A}}
\newcommand{\bfE}{{\bf E}}
\newcommand{\bfB}{{\bf B}}

\newcommand{\bfl}{{\bf l}}
\newcommand{\bfF}{{\bf F}}
\newcommand{\bfL}{{\bf L}}
\newcommand{\bfT}{{\bf T}}

\begin{document}


\title{Local conservative regularizations of compressible MHD and neutral flows}

\author{Govind S. Krishnaswami} 

\affiliation{Physics Department, Chennai Mathematical Institute,  SIPCOT IT Park, Siruseri 603103, India}

\email{govind@cmi.ac.in, sonakshi@cmi.ac.in}

\author{Sonakshi Sachdev} 

\affiliation{Physics Department, Chennai Mathematical Institute,  SIPCOT IT Park, Siruseri 603103, India}

\author{A. Thyagaraja}

\affiliation{Astrophysics Group, University of Bristol, Bristol, BS8 1TL, UK}

\email{athyagaraja@gmail.com}

\date{24 February, 2016, \quad Published in Phys. Plasmas 23, 022308 (2016).}

\begin{abstract} \normalsize

Ideal systems like MHD and Euler flow may develop singularities in vorticity ($\bfw = \grad \times \bfv$). Viscosity and resistivity provide dissipative regularizations of the singularities. In this paper we propose a minimal, local, conservative, nonlinear, dispersive regularization of compressible flow and ideal MHD, in analogy with the KdV regularization of the 1D kinematic wave equation. This work extends and significantly generalizes earlier work on incompressible Euler and ideal MHD. It involves a micro-scale cutoff length $\lambda$ which is a function of density, unlike in the incompressible case. In MHD, it can be taken to be of order the electron collisionless skin depth $c/\omega_{pe}$. Our regularization preserves the symmetries of the original systems, and with appropriate boundary conditions, leads to associated conservation laws. Energy and enstrophy are subject to a priori bounds determined by initial data in contrast to the unregularized systems. A Hamiltonian and Poisson bracket formulation is developed and applied to generalize the constitutive relation to bound higher moments of vorticity. A `swirl' velocity field is identified, and shown to transport $\bfw/\rho$ and $\bfB/\rho$, generalizing the Kelvin-Helmholtz and Alfv\'en theorems. The steady regularized equations are used to model a rotating vortex, MHD pinch and a plane vortex sheet. The proposed regularization could facilitate numerical simulations of fluid/MHD equations and provide a consistent statistical mechanics of vortices/current filaments in 3D, without blowup of enstrophy. Implications for detailed analyses of fluid and plasma dynamic systems arising from our work are briefly discussed.

\end{abstract}



\maketitle


\section{Introduction}

Compressible inviscid gas dynamics and ideal magnetohydrodynamics (MHD) have been remarkably fruitful areas of investigation and are relevant to modern aerodynamics, astrophysics and fusion plasma physics. It is well-known that time evolution of these ideal equations often leads to finite-time singularities or unbounded growth of vorticity and current (for a recent example see Ref.~\onlinecite{henneberg-cowley-wilson}). Apart from shock formation, which is relatively well-understood, the mechanism underlying singularities in these systems in three-dimensions (3D) is the phenomenon of vortex stretching [\onlinecite{Frisch},\onlinecite{KRSreenivasan-onsager}]. In an earlier work [\onlinecite{thyagaraja}], one of us
introduced local regularizing `twirl' terms ($\la^2 \bfw \times (\grad \times \bfw), \la^2 \bfB \times (\grad \times \bfw)$) in the inviscid incompressible Euler and MHD equations with the aim of guaranteeing an a priori upper bound on the enstrophy. This bound was determined by the initial data and the systems were shown to retain their conservation properties. The motivation for such a regularization is the observation that the KdV equation ($u_t - 6 u u_x + u_{xxx} = 0$) is a dispersive local regularization of the 1D kinematic wave equation (KWE: $u_t + u u_x = 0$), which is known to have finite-time singularities. It is well-known that the KdV equation is a canonical model of non-linear dispersive waves with applications in widely disparate fields [\onlinecite{Miura}, \onlinecite{davidson}]. Furthermore KdV has a Hamiltonian structure and its initial value problem in an infinite domain is exactly soluble [\onlinecite{Miura}, \onlinecite{ablowitz-clarkson}].

The purpose of this work is to (1) develop a local conservative regularization for 3D {\it compressible} gas dynamics and ideal MHD, (2) motivate the physical criteria underlying the regularization, (3) derive a Hamiltonian and Poisson structure and present the conservation laws implied by them and (4) present some simple solutions exemplifying new features of the regularized systems.

Our principal aim is to deduce that with the twirl regularization, our systems are both Hamiltonian and possess global upper bounds for enstrophy, kinetic and compressional energies determined by the initial data, guaranteeing `Lagrange' stability  of the system motion [\onlinecite{Nem}].

Our regularization procedure extending the earlier work [\onlinecite{thyagaraja}] is in the spirit of effective local field theory, typified by the short-range repulsive Skyrme term which stabilizes the singularity in the soliton solution of the QCD effective chiral Lagrangian [\onlinecite{bal-book}]. The regularization must respect global symmetries of the original system and possess corresponding local conservation laws. The added terms must be local and minimal in nonlinearity and derivatives and be `small' to leave the macro- and meso-scale dynamics unaltered. KdV certainly satisfies the above criteria, but involves a third order {\it linear} dispersive term. In marked contrast, our twirl terms are quadratically non-linear (important in high-speed compressible flows) and second order in velocity derivatives.

It is well-known that KWE admits Burgers' dissipative regularization $\nu u_{xx}$. Also well-known is the Navier-Stokes (NS) viscous regularization ($\nu \grad^2 \bfv$) of inviscid Euler equations with its counterpart in visco-resistive MHD. At present, it is an open problem whether these 3D dissipative systems are truly regular (i.e. with classical solutions for all $t > 0$). The situation is well reviewed in [\onlinecite{Frisch}, \onlinecite{KRSreenivasan-onsager}]. In particular, the existence of classical solutions has been shown by Ladyzhenskaya in Ref.~\onlinecite{ladyzhenskaya}, provided $\nu$ is not constant as in NS, but involves velocity gradients to some positive power (`hyperviscosity'). There also exist other regularizations of NS (NS-$\alpha$) based on non-local averaging of the advecting velocity, for which a proof of global regularity is available in Ref.~\onlinecite{foias-hom-titi}. As far as we are aware, these results pertain only to incompressible hydrodynamics and do not apply to compressible flows or visco-resistive MHD. We note that the hyperviscosity regulator, although dissipative, is non-linear and serves to balance, in principle, the non-linear vortex-stretching mechanism of 3D inviscid flow. Our non-linear twirl term is similarly responsible for controlling the growth of enstrophy at short distances of order $\la$ (as demonstrated in the incompressible case in Ref.~\onlinecite{thyagaraja}). A crucial difference is that like KdV, our models are both conservative and local, unlike hyperviscosity and NS-$\alpha$ models which are dissipative and pertain to driven systems.

A key feature of our twirl regularization is the introduction of a length scale $\la \ll L$, where $L$ is a macroscopic length. $\la$ plays the role of the Taylor length [\onlinecite{Frisch}] in hydrodynamics and the electron collisionless skin depth $\delta = c/\om_{pe}$ in MHD. Thus the twirl term acts as a short-distance cutoff preventing excessive production of enstrophy at that length scale. In the incompressible case, $\la$ was a constant. In the compressible models, it must satisfy a constitutive relation $\la^2 \rho =$ constant. This relation might be expected from the following vortical-magnetic analogy. Indeed, the twirl force $- \la^2 \rho \, \bfw \times (\grad \times \bfw)$ is a vortical counterpart of the magnetic Lorentz force $\bfj \times \bfB = -  [\bfB \times (\grad \times \bfB)]/\mu_0$, with $\la^2 \rho$ replacing the constant $1/\mu_0$. The constitutive relation can be interpreted in terms of the mean free path   and/or the inter-particle distance $a \propto n^{-1/3}$ where $n$ is the number density of molecules in the medium [\onlinecite{thyagaraja}]. Thus if we take $(\la/L)^2 \propto (a/L)^3$ then $\la^2 \rho$ will be a constant. In plasmas there are natural length-scales inversely proportional to the square-root of the number density. For example, the skin-depth $\del \propto 1/\sqrt{n_e}$. Thus if $\la \approx\del$ then $\la^2 \rho$ will be a constant. In any event, it is well-known that ideal MHD is not valid at length scales of order $\del$. Another example is provided by the electron Debye length $\la_D = \sqrt{2 T_e \eps_0/n_e e^2}$ in an isothermal plasma. If we take $\la/\la_D$ constant, then we recover the postulated constitutive relation. Thus, having a cut-off of this kind will not affect any major consequence of ideal MHD on meso- and macro-scales and yet provide an upper bound to the enstrophy of the system. 

Inclusion of these twirl regularizations should lead to more controlled numerical simulations of Euler, NS and MHD equations without finite time blowups of enstrophy. In particular, these regularized models are capable of handling 3D tangled vortex line and sheet interactions in engineering and geophysical fluid flows, corresponding current filament and sheet dynamics which occur in astrophysics (e.g. as in solar prominences and coronal mass ejections, pulsar accretion disks and associated turbulent jets, and on a galactic scale, jets driven by active galactic nuclei) as well as in strongly nonlinear phenomena such as edge localised modes in tokamaks. There is no known way of studying many of these phenomena at very low collisionality  [i.e. at very high, experimentally relevant Reynolds, Mach and Lundquist numbers] with unregularized continuum models. Thus, we  note that recent theories [\onlinecite{henneberg-cowley-wilson}, \onlinecite{chandra-thyagaraja}, \onlinecite{lashmore-mccarthy-thyagaraja}, \onlinecite{thyagaraja-valovic-knight}] of the nonlinear evolution of ideal and visco-resistive plasma turbulence in a variety of fusion-relevant devices (and many geophysical situations) can be numerically investigated in a practical way using our regularization.

The existence of a positive definite Hamiltonian and bounded enstrophy should facilitate the formulation of a valid statistical mechanics of 3D vortex tubes, extending the work of Onsager 
[\onlinecite{KRSreenivasan-onsager}] on 2D line vortices. The same applies to the possible extension of 2D statistical mechanics of line current filaments developed by Edwards and Taylor in Ref.~\onlinecite{edwards-taylor} in incompressible ideal MHD.

We begin in \S\ref{s:formulation} by formulating the equations of regularized compressible flow and MHD. The nature of the quadratically non-linear twirl term and the constitutive relation are discussed. In \S\ref{s:cons-laws} local conservation laws, boundary conditions, global integral invariants obtained from them and freezing-in theorems generalizing Kelvin-Helmholtz and Alfv\'en are derived. Integral invariants associated to the `swirl' velocity are discussed in \S\ref{s:swirl-vel-and-intergral-invariants}. A Hamiltonian formulation based on the elegant Landau-Morrison-Greene [\onlinecite{landau}, \onlinecite{morrison-and-greene}] Poisson brackets is presented in \S\ref{s:pb-for-fluid}. It is used to identify new conservative regularizations that guarantee bounded higher moments of vorticity. \S \ref{s:examples} contains applications to regularized steady flows in a magnetized columnar vortex/MHD pinch and a vortex sheet. Conclusions are presented in \S\ref{s:discussion}. 

\vspace{-.5cm}

\section{Formulation of regularized models}
\label{s:formulation}

For compressible flow with mass density $\rho$ and velocity field $\bfv$, the continuity and Euler equations are
	\beq
	\dd{\rho}{t} + \grad \cdot (\rho {\bf v}) = 0 \quad \text{and} \quad
	\dd{\bf v}{t} + \left( {\bf v} \cdot \grad \right) {\bf v} = - \frac{\grad  p}{\rho}.
	\label{e:Euler-eqn-compressible}
	\eeq
The pressure $p$ is related to $\rho$ through a constitutive relation in barotropic flow. The stagnation pressure $\sigma$ and specific enthalpy $h$ for adiabatic flow of an ideal gas are
	\beq
	\sigma \equiv h + \frac{\bfv^2}{2} = \left( \frac{\gamma}{\gamma -1} \right) \frac{p}{\rho} + \frac{{\bf v}^2}{2} 
	\eeq
where $p/\rho^\gamma$ is constant, with $\gamma = C_p/C_v$. Then using the identity $\half \grad {\bf v}^2 = {\bf v} \times \left( \grad \times {\bf v} \right) + \left({\bf v} \cdot \grad \right) {\bf v}$, the Euler equation may be written in terms of vorticity $\bfw = \grad \times \bfv$;
	\beq
	\dd{\bf v}{t} + {\bf w} \times {\bf v} = - \grad \sigma.
	\label{e:Euler-eqn}
	\eeq
In Ref.~\onlinecite{thyagaraja} a regularizing twirl acceleration term $- \la^2 \bfT$ was introduced in the incompressible $(\grad \cdot \bfv = 0)$ Euler equation
	\beq
	\dd{\bf v}{t} + \left( {\bf v} \cdot \grad \right) {\bf v} = - \frac{\grad p}{\rho} - \la^2 {\bf w} \times (\grad \times {\bf w}). 
	\label{e:R-Euler-eqn}
	\eeq
The twirl term is a singular perturbation, making the regularized Euler (R-Euler) equation $2^{\rm nd}$ order in space derivatives of $\bfv$ while remaining $1^{\rm st}$ order in time. The parameter $\la$ with dimensions of length is a constant for incompressible flow. The twirl term $- \la^2 \bfT$ is a conservative analogue of the viscous dissipation term $\nu \grad^2 \bfv$ in the incompressible NS equation
	\beq
	\dd{\bfv}{t} + (\bfv \cdot \grad) \bfv = - \frac{\grad p}{\rho} + \nu \grad^2 \bfv, \quad \grad \cdot \bfv = 0.
	\eeq
Kinematic viscosity $\nu$ and the regulator $\la$ play similar roles. The momentum diffusive time scale in NS is set by $\nu k^2$ where $k$ is the wave number of a mode. On the other hand in the non-linear twirl term of R-Euler, the dispersion time-scale of momentum is set by $\la^2 k^2 |\bfw|$. So for high vorticity short wavelength modes, the twirl effect would be more efficient in controlling enstrophy than pure viscous diffusion. 

It is instructive to compare the relative sizes of the dissipative stress and the conservative twirl force in vorticity equations. Under the usual rescaling $\bfr = L \bfr', \bfv = U \bfv'$ ($t = (L/U) t'$, $\bfw = (U/L) \bfw'$) and $|\grad| = k$, $F_{visc} \sim (\nu/L^2) k^2 \om$ whereas $F_{twirl} \sim (\la^2 U/L^3) k^2 \om^2$ where $\om$ is the magnitude of the non-dimensional vorticity. Then $F_{twirl}/F_{visc} \sim {\cal R} \om (\la/L)^2$. This shows that at any given Reynolds number ${\cal R} = LU/\nu$ and however small $\la/L$ is taken, at sufficiently large vorticity the twirl force will always be larger than the viscous force.

It is also interesting to compare incompressible Euler, NS and R-Euler under rescaling of coordinates $\bfr = L \bfr'$ and velocities $\bfv = U \bfv'$ . The incompressible Euler equations for vorticity
are invariant under such rescalings. The NS equation is not invariant unless $LU = 1$.
Interestingly, the incompressible R-Euler equation for vorticity is invariant under rescaling of time but not space, due to the presence of the length scale $\la$.

Since $\bfT$ is quadratic in $\bfv$, it should be important in high-speed flows as in compressible gas dynamics. Consider adiabatic flow of an ideal fluid with adiabatic equation of state: $(p/p_0) = (\rho/\rho_0)^\gamma$. The {\it compressible} R-Euler momentum equation is
	\beq
	\dd{\bf v}{t} + \left( {\bf v} \cdot \grad \right) {\bf v} = - \grad h - \la^2 {\bf w} \times (\grad \times {\bf w}).
	\label{e:R-Euler-compressible}
	\eeq
To ensure that a positive-definite conserved energy exists for an arbitrary flow [more general constitutive relations are derived in \S\ref{s:pb-for-fluid}] we find that $\la(\bfr,t)$ and $\rho(\bfr,t)$ must satisfy a constitutive relation (to be discussed shortly):
	\beq
	\la^2 \rho = \text{constant} = \la_0^2  \rho_0.
	\label{e:constitutive-relation}
	\eeq
The constant $\la_0^2 \rho_0$ is a property of the fluid, like kinematic viscosity. As before, we write R-Euler as
	\beq
	\bfv_t + \bfw \times \bfv = - \grad \sigma - \la^2 \bfw \times (\grad \times \bfw).
	\label{e:reg-Euler-3D}
	\eeq
Here $\bfw \times \bfv$ is the `vorticity acceleration' and $- \la^2 \bfw \times (\grad \times \bfw)$ is the twirl acceleration while $\grad \sigma$ includes acceleration due to pressure gradients. The regularization term increases the spatial order of the Euler equation by one, just as  $\nu \grad^2 \bfv$ in going from Euler to NS. However, the boundary conditions (see \S\ref{s:cons-laws}) required by the above conservative regularization involve the first spatial derivatives of $\bfv$, unlike the no-slip condition of NS. Both the twirl and dispersive term in KdV involve three derivatives of velocity; however, the former is second order and quadratic unlike the 3rd order and linear $u_{xxx}$ term of KdV. It should be noted that the vortex-stretching inertial term of the Euler equation is balanced by a linear diffusion term in NS whereas it is balanced by a quadratically non-linear dispersive twirl term in R-Euler. The R-Euler equation is invariant under parity (all terms reverse sign) and under time-reversal. It is well-known that NS is not invariant under time-reversal, since it includes viscous dissipation. The R-Euler equation takes a compact form in terms of the {\it swirl} velocity field $\bfv_* = \bfv + \la^2 \grad \times \bfw$:
	\beq
	\dd{\bfv}{t} + \bfw \times \bfv_* = - \grad \sigma.
	\label{e:R-euler-v*-sigma}
	\eeq
Note that $\bfv_*$ differs little from $\bfv$ on length-scales large compared to $\la$. $\bfw \times \bfv_*$ is a regularized version of the Eulerian vorticity acceleration $\bfw \times \bfv$. The swirl velocity plays an important role in the regularized theory, as will be demonstrated. In fact, the continuity equation can be written in terms of $\bfv_*$ using the constitutive relation  (\ref{e:constitutive-relation}) 
	\beq
	\dd{\rho}{t} + \grad \cdot (\rho \bfv_*) = 0.
	\label{e:v*-continuity-eqn}
	\eeq
Taking the curl of (\ref{e:R-euler-v*-sigma}) we get the R-vorticity equation:
	\beq
	\bfw_t + \grad \times (\bfw \times \bfv_*) = 0.
	\label{e:R-vorticity-eqn-compressible}
	\eeq
With suitable boundary data, the incompressible regularized evolution equations possess a positive definite integral invariant (`swirl' energy in flow domain $V$):
	\beq
	\DD{E^*}{t} = \DD{}{t} \int_V \left[ \half \rho {\bf v}^2 + \half \la^2 \rho {\bf w}^2 \right] \: d\bfr = 0.
	\label{e:cons-egy-incompress}
	\eeq
For compressible flow, $E^*$ is {\em not} conserved if $\la$ is a constant length. On the other hand, we do find a conserved swirl energy if we include compressional potential energy and also let $\la(\bfr,t)$ be a dynamical length governed by the constitutive relation (\ref{e:constitutive-relation}). Here $\la_0$ is some constant short-distance cut-off (e.g. a mean-free path at mean density) and $\rho_0$ is a constant mass density (e.g. the mean density). $\la$ is smaller where the fluid is denser and larger where it is rarer. This is reasonable if we think of $\la$ as a position-dependent mean-free-path. However, it is only the combination $\la_0^2 \:\rho_0$ that appears in the equations. So compressible R-Euler involves only one new dimensional parameter, say $\la_0$. A dimensionless measure of the cutoff $n \la^3 = \la_0^3 \, n_0^{3/2} \,n^{-1/2}$ may be obtained by introducing the number density $n = \rho/m$ where $m$ is the molecular mass. It is clearly smaller in denser regions and larger in rarified regions. The R-Euler system is readily extended to include conservative body forces $\bfF = - \rho \grad V$ by adding $V$ to $\sigma$. This extension would be relevant for gravitational systems encountered in astrophysics. 

A much less trivial extension is to compressible ideal MHD. It is well-known that the governing equations for a quasi-neutral barotropic compressible ideal magnetized fluid [\onlinecite{hazeltine-meiss}] are 
	\beqs
	\rho_t + \grad \cdot (\rho \bfv) &=& 0, \quad
	\bfv_t + (\bfv \cdot \grad ) \bfv = - \ov{\rho} \grad p + \frac{\bfj \times \bfB}{\rho}
	\cr
	\text{and} \quad
	\bfB_t &=& \grad \times (\bfv \times \bfB)
	\label{e:unreg-idela-MHD}
	\eeqs
where $\mu_0 \bfj = \grad \times \bfB$. As usual, the electric field is given by the MHD Ohm's law, $\bfE + \bfv \times \bfB = 0$.

The regularized compressible MHD (R-MHD) equations follow from arguments similar to those for neutral compressible flows. The continuity equation (\ref{e:v*-continuity-eqn}), $\rho_t + \grad \cdot (\rho \bfv) = 0$ is unchanged. $\la$ is again subject to (\ref{e:constitutive-relation}). Thus (\ref{e:v*-continuity-eqn}) may be written in terms of swirl velocity: $\rho_t + \grad \cdot (\rho \bfv_*) = 0$. As in regularized fluid theory, we introduce the twirl acceleration on the RHS in the momentum equation, 
	\beq
	\bfv_t + \bfv \cdot \grad \bfv 
	= -\frac{\grad p}{\rho} - \frac{\bfB \times(\grad \times\bfB)}{\mu_0\rho} - \la^2 \bfw \times (\grad \times \bfw).
	\label{e:R-MHD-mom-eqn}
	\eeq

Eq. (\ref{e:R-MHD-mom-eqn}) can be written in terms of $\bfv_*$ as in R-Euler:
	\beq
	\dd{\bfv}{t} + \bfw \times \bfv_* = -\ov{\rho}\grad p - \half \grad \bfv^2 + \frac{\bfj \times \bfB}{\rho}.
	\label{e:R-MHD-Euler}
	\eeq
Faraday's law is regularized by replacing $\bfv$ by $\bfv_*$:
	\beq
	\pdr_t \bfB = \grad \times (\bfv_* \times \bfB)		\label{e:R-MHD-Faraday}
	\eeq
As in ideal MHD, the evolution equations for $\bfB$ and $\bfw$ (\ref{e:R-vorticity-eqn-compressible}) have the same form. Important physical consequences of this will be discussed in \S \ref{s:swirl-vel-and-intergral-invariants}. The regularization term in Faraday's law is the curl of the `magnetic' twirl term $- \la^2 \bfB \times (\grad \times \bfw)$ in analogy with the `vortical' twirl term $-\la^2 \bfw \times (\grad \times \bfw)$. The regularized Faraday law is $3^{\rm rd}$ order in space derivatives of $\bfv$ (as is the R-vorticity equation) and first order in $\bfB$. From (\ref{e:R-MHD-Faraday}) we deduce that the potentials ($\bfA, \phi$) in any gauge must satisfy
	\beq
	\pdr_t \bfA = \bfv_* \times \bfB - \grad \phi.	
	\label{e:eom-for-A-and-phi-R-MHD}
	\eeq
As before, conservative body forces like gravity are readily included in R-MHD as would be required in the dynamics of pulsar accretion disks. The inclusion of $\kappa$ and $\mu$ terms of Ref.~[\onlinecite{thyagaraja}] associated with electron inertia and Hall effect will be considered in a later work.

\section{Conservation laws}
\label{s:cons-laws}

{\noindent \bf Swirl Energy:} Under compressible R-Euler evolution, the swirl energy density and flux vector
	\beqs
	{\cal E}^* &=& \half {\rho \bfv^2} + U(\rho) + \half {\la^2 \rho \bfw^2}
	\quad \text{and} \quad
	\cr
	\bff &=& \rho \sigma \bfv + \la^2 \rho (\bfw \times \bfv) \times \bfw + \la^4 \rho \: \bfT \times \bfw
	\label{e:swirl-egy-density-current}
	\eeqs
satisfy the local conservation law $\pdr_t {\cal E}^* + \grad \cdot \bff = 0$. Here $U(\rho) = p/(\gamma - 1)$ is the compressional potential energy for adiabatic flow. Given suitable boundary conditions [BCs, see below], the system obeys a global energy conservation law $\dot E^* = 0$ where
	\beq
	E^* = \int \left[ \frac{\rho \bfv^2}{2}+ U(\rho) + \frac{\la^2 \rho \bfw^2}{2} \right] \, d\bfr.
	\label{e:swirl-energy-R-Euler}
	\eeq
{\noindent \bf Flow Helicity:} Compressible R-Euler equations possess a locally conserved helicity density $\bfv \cdot \bfw$ and flux $\bff_{\cal K}$:
	\beq
	\pdr_t (\bfv \cdot \bfw)
	+ \grad \cdot \left( \sig \bfw + \bfv \times (\bfv \times \bfw) + \la^2 {\bf T} \times \bfv \right) = 0.
	\label{e:helicity-current-conservation}
	\eeq
If $\bff_{\cal K} \cdot \hat n  = 0$ on the boundary $\pdr V$ of the flow domain, then helicity ${\cal K} = \int \bfv \cdot \bfw \, d\bfr$ is a constant of motion.

{\noindent \bf Momentum:} Flow momentum density ${\cal P}_i = \rho v_i$ and the stress tensor $\Pi_{ij}$ satisfy $\pdr_t {\cal P}_i + \pdr_j \Pi_{ij} = 0$ where
	\beq
	\Pi_{ij} = \rho v_i v_j + p \del_{ij} + \la^2 \rho \left(\half \bfw^2 \del_{ij} - w_i w_j \right).
	\label{e:momentum-current-tensor}
	\eeq
For ${\bf P} = \int \rho \bfv \; d\bfr$ to be conserved, we expect to need a translation-invariant flow domain $V$. If $V = \mathbb{R}^3$, decaying boundary conditions ($\bfv \to 0$) ensure $\dot{\bf P} = 0$. Periodic BCs in a cuboid also ensure global conservation of $\bf P$.

{\noindent \bf Angular momentum:} For regularized compressible flow, angular momentum density $\vec {\cal L} = \rho {\bf r} \times \bfv$ and its current tensor $\Lambda$ satisfy the local conservation law:
	\beq
	\pdr_t {\cal L}_i + \pdr_l \Lambda_{il} = 0 \quad \text{where} \quad
	\Lambda_{il} = \eps_{ijk} r_j \Pi_{kl}.
	\label{e:ang-mom-current-conservation}
	\eeq
For $\bfL = \int \vec {\cal L} \: d\bfr$ to be globally conserved, the system must be rotationally invariant. For instance, decaying BCs in $\mathbb{R}^3$ would guarantee conservation of $\vec {\cal L}$. In symmetric domains [axisymmetric torus or circular cylinder] corresponding components of $\bfL$ or $\bfP$ may also be conserved. 

The proofs of these local conservation laws follow from the equations of motion and constitutive relation (\ref{e:constitutive-relation}) as described in Ref.~\onlinecite{thyagaraja} for incompressible flow. Details may be found in Ref.~\onlinecite{arxiv-version}. In \S \ref{s:pb-for-fluid} we show that these conservation laws and boundary conditions also follow from a Hamiltonian/Poisson bracket formulation.

{\flushleft \bf Boundary conditions:} In the flow domain $\mathbb{R}^3$, it is natural to impose decaying BCs ($\bfv \to 0$ and $\rho \to$ constant as $|\bfr| \to \infty$) to ensure that total energy $E^*$ is finite and conserved. For flow in a cuboid, periodic BCs ensure finiteness and conservation of energy. For flow in a bounded domain $V$, demanding global conservation of energy leads to another natural set of BCs. Now $\dot E^* = - \int_{\pdr V} \bff \cdot \hat n \: dS$ where $\bff$ is the energy current (\ref{e:swirl-egy-density-current}) and $\pdr V$ the boundary surface. The flux $\bff \cdot \hat n = 0$ if
	\beq
	\bfv \cdot \hat n = 0 \quad
	\text{and} \quad
	\bfw \times \hat n = 0.
	\label{e:BCs-for-energy-cons}
	\eeq
As R-Euler is $2^{\rm nd}$ order in spatial derivatives of $\bfv$, it is consistent to impose BCs on $\bfv$ and its $1^{\rm st}$ derivatives. These BCs imply that the twirl acceleration is tangential to $\pdr V$: $\bfT \cdot \hat n = 0$.

Interestingly, BCs ensuring helicity conservation are `orthogonal' to those for $E^*$ conservation
	\beq
	\bfv \times \hat n = 0 \quad 
	\text{and} \quad
	\bfw \cdot \hat n = 0 \quad \imply \quad \bff_{\cal K} \cdot \hat n = 0.
	\eeq
However, periodic or decaying BCs would ensure simultaneous conservation of both $E^*$ and $\cal K$. 
{\flushleft \bf R-MHD swirl energy:} In barotropic ($\grad U'(\rho) = \grad p/\rho$) compressible R-MHD, swirl energy is locally conserved: $\pdr_t {\cal E}^* + \grad \cdot \bff = 0$ where
	\beqs
	&& {\cal E}^* = \frac{\rho \bfv^2}{2}+ U(\rho) + \frac{\la^2 \rho \bfw^2}{2} + \frac{\bfB^2}{2 \mu_0} 
	\quad  \rm{and} 
	\cr 
	&& \bff 
	= \rho \sigma \bfv + \la^2 \rho (\bfw \times \bfv) \times \bfw + \la^4 \rho \bfT \times \bfw + \frac{\bfB \times (\bfv \times \bfB)}{\mu_0} \cr
	&& + \frac{\la^2}{\mu_0} \left\{ \bfw \times ((\grad \times \bfB) \times \bfB)
	+ \bfB \times ( (\grad \times \bfw ) \times \bfB )\right\}
	\eeqs
is the energy flux vector and $E^*_{\rm mhd} = \int_V{\cal E}^* \: d^3r$ is the the total `swirl' energy. BCs that ensure  global conservation of $E^*_{\rm mhd}$ follow from requiring $\bff \cdot \hat n = 0$: \beq
	\bfv \cdot \hat n = 0, \;\;
	\bfw \times \hat n = 0, \;\; 
	\grad \times \bfw \, \cdot \hat n = 0 \;\; \text{and} \;\;
	\bfB \cdot \hat n = 0.
	\label{e:R-MHD-energy-BCs}
	\eeq
The R-MHD equations (\ref{e:R-MHD-Euler},\ref{e:R-MHD-Faraday}) are $3^{\rm rd}$ order in $\bfv$ and $1^{\rm st}$ order in $\bfB$. So we may impose BCs on $\bfB$, $\bfv$, the $1^{\rm st}$ and $2^{\rm nd}$ derivatives of $\bfv$. (\ref{e:R-MHD-energy-BCs}) also implies that $\bfB \cdot \bfw = 0$ and $\bfv_* \cdot \hat n = 0$ on the boundary. For details see Ref.~\onlinecite{arxiv-version}.

The conservation of $E^*_{\rm mhd}$ implies a global a priori bound on kinetic and magnetic energies, and most importantly, enstrophy. Such a bound on enstrophy is not available for compressible Euler or ideal MHD. This has important physical consequences which will be discussed.

{\flushleft \bf Magnetic helicity:} Magnetic helicity ${\cal K}_{B} = \int_V \bfA \cdot \bfB$ is the magnetic analogue of flow helicity. Its density and current are locally conserved in R-MHD. Using (\ref{e:R-MHD-Faraday}, \ref{e:eom-for-A-and-phi-R-MHD}) 
	\beq
	\pdr_t (\bfA \cdot \bfB) + \grad \cdot (\bfA \times (\bfv_* \times \bfB) + \bfB \phi) = 0.
	\label{e:mag-helicity-and-current-conservation}
	\eeq
Global conservation of ${\cal K}_{B}$ requires vanishing flux of magnetic helicity across $\pdr V$. This is guaranteed if $\bfB \cdot \hat n = 0$, $\bfv \cdot \hat n = 0$ and $(\grad \times \bfw) \cdot \hat n = 0$. Note that for conservation of ${\cal K}_{B}$ it suffices that both $\bfB$ and $\bfv_*$ be tangential to $\pdr V$. $\bfB \cdot \hat n = 0$ also ensures gauge-invariance of ${\cal K}_B$. Unlike for flow helicity, the BCs that guarantee $E_{\rm mhd}^*$ conservation also ensure conservation of ${\cal K}_{B}$ (though not vice versa). 

{\noindent \bf Linear and angular momenta in R-MHD:} The momentum density ${\cal P}_i = \rho v_i$ and stress tensor $\Pi_{ij}$ satisfy a local conservation law $\pdr_t {\cal P}_i + \pdr_j \Pi_{ij} = 0$ where $\Pi_{ij}$ is
	\beq
	\rho v_i v_j + p \del_{ij} + \la^2 \rho \left[ \frac{\bfw^2}{2} \del_{ij} - w_i w_j \right] + \frac{\bfB^2}{2\mu_0} \del_{ij} - \frac{B_i B_j}{\mu_0}.
	\eeq
$\bfB$ and $\bfw$ enter $\Pi_{ij}$ in the same manner as the twirl force $-\la^2 \rho \bfw \times (\grad \times \bfw)$ and magnetic Lorentz force $- (\bfB \times (\grad \times \bfB))/\mu_0$ are of the same form. We define angular momentum density in R-MHD as $\vec {\cal L} = \rho \bfr \times \bfv$. Using the local conservation of $\rho \bfv$ we find that $\vec {\cal L}$ too is locally conserved in R-MHD:
	\beq
	\pdr_t {\cal L}_i 
	= - \pdr_l \left( \eps_{ijk} r_j \Pi_{kl} \right) = - \pdr_l \Lambda_{il}.
	\eeq
Total momentum $\int {\cal P}_i \, d\bfr$ and angular momentum $\int {\cal L}_i \, d\bfr$ are globally conserved for appropriate boundary conditions (e.g. decaying BC in an infinite domain or periodic BC in a cuboid for momentum).

{\flushleft \bf Swirl velocity and `freezing-in' theorems in R-Euler and R-MHD:} We note several interesting properties of swirl velocity $\bfv_* = \bfv + \la^2 \grad \times \bfw$ and its role in obtaining analogues of the well-known  Kelvin-Helmholtz and Alf\'ven theorems in R-Euler and R-MHD [\onlinecite{lamb}, \onlinecite{hazeltine-meiss}]. For instance, $\bfw/\rho$ is frozen into $\bfv_*$ (but not $\bfv$). The R-vorticity (\ref{e:R-vorticity-eqn-compressible}) and continuity (\ref{e:v*-continuity-eqn}) equations imply
	\beq
	\pdr_t (\bfw/\rho) + (\bfv_* \cdot \grad) (\bfw/\rho) = ((\bfw/\rho) \cdot \grad) \bfv_* 
	\label{e:wbyrho-frozen-in-v*}
	\eeq
Similarly, (\ref{e:R-MHD-Faraday}) and continuity equation(\ref{e:v*-continuity-eqn}) implies that $\bfB/\rho$ is frozen into $\bfv_*$:
	\beq
	\pdr_t (\bfB/\rho) + (\bfv_* \cdot \grad) (\bfB/\rho) = \left( (\bfB/\rho) \cdot \grad \right) \bfv_*.
	\label{e:Bbyrho-frozen-in-v*}
	\eeq


{\noindent \bf Swirl energy in terms of $\bfv_*$:} In both R-Euler and R-MHD, $E^*$ is expressible in terms of $\bfv_*$. Up to a boundary term (which vanishes if $\bfv \times \hat n = 0$ or $\bfw \times \hat n = 0$), $\bfv \cdot \bfv_*$ accounts for both kinetic and enstrophic energies:
	\beq
	E^* = \int_V \left(\ov{2}\rho(\bfr)\bfv_*(\bfr)\cdot \bfv(\bfr) + U(\rho) + \frac{\bfB^2(\bfr)}{2 \mu_0} \right) \: d\bfr.
	\eeq

\section{Swirl velocity and integral invariants}
\label{s:swirl-vel-and-intergral-invariants}

{\noindent \bf Swirl Kelvin circulation theorem:} The circulation $\G$ of $\bfv$ around a closed contour $C^*$ (that moves with $\bfv_*$) is independent of time. This is a regularized version of the Kelvin circulation theorem.
	\beq
		\frac{d\G}{dt} = \frac{d}{dt} \oint_{C^*} \bfv \cdot d\bfl = \frac{d}{dt} \int_{S^*} \bfw \cdot d\bfS = 0.
	\label{e:swirl-kelvin-theorem}
	\eeq
The second equality follows by Stokes' theorem. Here $S^*$ is any surface spanning $C^*$. 
Consider
	\beq
	\DD{}{t} \oint_{C^*} \bfv \cdot d\bfl 
	= \oint_{C^*} \left( \dd{\bfv}{t} + \bfv_* \cdot \grad \bfv \right) \cdot d\bfl + \oint_{C^*} \bfv \cdot d\bfv_*.
	\eeq
Noting that $D_t^* = \pdr_t + \bfv_* \cdot \grad$, as the line element moves with $\bfv_*$, $D_t^*(d\bfl) = d (D_t^* \bfl) = d \bfv_*$. From (\ref{e:R-euler-v*-sigma}) and the identity $\bfv_* \cdot \grad \bfv =   \grad \bfv \cdot \bfv_*- \bfv_* \times (\grad \times \bfv)$ we get
	\beq
	\dot \G = \oint_{C^*}  \left[(\grad \bfv \cdot \bfv_* - \grad \sigma) \cdot d\bfl + \bfv \cdot d \bfv_*\right]
	= \oint_{C^*} d (\bfv_* \cdot \bfv) = 0.
	\eeq

{\flushleft \bf Swirl Alfv\'en theorem on magnetic flux:} The line integral of the magnetic vector potential $\Phi = \oint_{C^*} \bfA \cdot d\bfl$ over a closed contour $C^*_t$ moving with $\bfv_*$ is a constant of the motion. The proof is similar to that of the swirl Kelvin theorem and uses the equation of motion $\pdr\bfA/\pdr t = \bfv_* \times \bfB - \grad \phi$. Now if $S^*$ is any surface spanning the contour $C^*$ then from Stokes' theorem we see that $\Phi = \int_{S^*} \bfB \cdot d\bfS$ is a constant of the motion. This is the regularized version of Alfv\'en's frozen-in flux theorem.

A smooth function $S(\bfx,t)$ which satisfies $D^*_t S \equiv \pdr_t S + {\bf v}_* \cdot \grad S = 0$ defines (level) surfaces which move with $\bfv_*$. They enclose volumes $V^*$ that move with $\bfv_*$. The freezing of $\bfB/\rho$ into $\bfv_*$ in R-MHD implies that $D^*_t((\bfB/\rho) \cdot \grad S) = 0$. The same holds for $\bfw/\rho$ in R-Euler. Thus magnetic flux tubes and vortex tubes move with $\bfv_*$.

{\flushleft \bf Constancy of mass of fluid in a volume $V^*$ moving with $\bfv_*$:} This follows by taking $f= \rho$ in the $\bfv_*$ analogue of the Reynolds' transport theorem
	\beq
	\frac{d}{dt}\int_{V^*} f \, d\bfr =\int_{V^*} D^*_t \left(\frac{f}{\rho}\right)\rho \: d \bfr.
	\label{e:reynolds-transport-theorem}
	\eeq

{\noindent \bf Conservation of flow helicity in a closed vortex tube:} The flow helicity ${\cal K}$ associated with a vortex tube enclosing a volume $V^*$ is independent of time:
	\beq
	\dot {\cal K} = \frac{d}{dt}\int_{V^*} \bfw \cdot \bfv \: d\bfr = 0.
	\eeq
Applying equations (\ref{e:reynolds-transport-theorem}), (\ref{e:wbyrho-frozen-in-v*}) and (\ref{e:R-euler-v*-sigma}) we get 
	\beqs
	\dot {\cal K} &=& \int_{V^*} \left[D^{*}_{t}\left(\frac{\bf w}{\rho}\right)\cdot{\bf v}+\left(\frac{\bf w}{\rho}\right)\cdot D^{*}_{t} ({\bf v})\right] \rho d\bfr, \cr
	&=& \int_{V^*} {\bf w}\cdot\left[ \grad\bfv_* \cdot \bfv + {\bf v}_{*}\cdot {\bf \grad v}+{\bf v}_{*}\times {\bf w} - \grad \sigma \right] d\bfr \cr
	&=& \int_{\pdr V^*} (\bfv \cdot \bfv_* - \sigma) \bfw \cdot \hat n \, dS = 0,
  \eeqs
since $\bfw$ is tangential to the vortex tube.

{\flushleft \bf Conservation of magnetic helicity in a magnetic flux tube} In R-MHD, the magnetic helicity ${\cal K}_{B}$ (but {\it not} flow helicity) in a volume bounded by a closed magnetic flux tube is independent of time:
	\beq
	\dot {\cal K}_{B} = \frac{d}{dt}\int_{V^*} \bfB \cdot \bfA \,d\bfx 	
	= \int_{\pdr V^*} (\bfA \cdot \bfv_* - \phi) \bfB \cdot \hat n dS = 0.
   	\eeq
This follows from Eqs. (\ref{e:reynolds-transport-theorem}), (\ref{e:Bbyrho-frozen-in-v*}) and (\ref{e:eom-for-A-and-phi-R-MHD}).

More generally, a Helmholtz field (see Ref.~\onlinecite{thyagaraja-IITM}) is a solenoidal field $\bfg$ evolving according to $\bfg_t + \grad \times (\bfg \times \bfv_*) = 0$. As above we deduce generalized Kelvin theorems for Helmholtz fields:
	\beqs
	&& \DD{}{t} \oint_{C^*} \bfu \cdot d\bfl = \DD{}{t} \int_{S^*} \bfg \cdot d\bfS = 0 \quad \text{and} \cr
	&& \DD{}{t} \int_{V^*} \bfg \cdot {\bf u} \: d\bfr = \int_{V^*} D_t^* (\bfg \cdot {\bf u}) \: d\bfr = 0
	\eeqs
where we evidently have the freezing-in equation $D_t^* (\bfg/\rho) = (\bfg/\rho) \cdot \grad \bfv_*$ and the `potential' equations 
	\beq
	\bfg = \grad \times {\bf u} \quad \text{with} \quad {\bf u}_t + \bfg \times \bfv_* + \grad \tht  = 0.
	\eeq
Here $V^*$ is the volume enclosed by a $\bfg$-tube, a closed surface everywhere tangent to $\bfg$. $C^*$ is a closed contour moving with $\bfv_*$ and $S^*$ is a surface spanning $C^*$. Examples of Helmholtz fields in R-Euler and R-MHD include $\bfw$ and $\bfB$. 
\section{Hamiltonian and Poisson structure}
\label{s:pb-for-fluid}

Commutation relations among `quantized' fluid variables were proposed by Landau in Ref.~\onlinecite{landau}. 
As a byproduct, one obtains Poisson brackets (PB) among {\it classical} fluid variables allowing a Hamiltonian formulation for compressible flow (due to Morrison and Greene [\onlinecite{morrison-and-greene}]). Suppose $F$ and $G$ are two functionals of $\rho$ and $\bfv$, then their equal-time PB is
	\beq
	\{ F, G \} = \int \left[\frac{\bfw}{\rho} \cdot F_\bfv \times G_\bfv - F_\bfv \cdot \grad G_\rho + F \leftrightarrow G \right]d\bfr
	\label{e:pb-between-functionals-of-rho-v}
	\eeq
where subscripts denote functional derivatives, e.g. $F_\rho = {\del F}/{\del \rho}$. The PB is manifestly anti-symmetric and has dimensions of $FG/\hbar$. This non-canonical PB satisfies the Leibnitz rule $\{ FG, H \} = F \{ G, H\} + \{F, H \} G$. 

From (\ref{e:pb-between-functionals-of-rho-v}) we deduce the PB among basic dynamical variables. Density commutes with itself $\{ \rho(\bfx), \rho(\bfy) \} = 0$, with $\la$ (c.f. (\ref{e:constitutive-relation})) and notably with vorticity. The non-trivial PBs are
	\beqs
	\{ v_i(\bfx), v_j(\bfy) \} &=& (\om_{ij}/\rho)(\bfx) \: \del(\bfx-\bfy),
	\cr
	\{ \rho(\bfx), \bfv(\bfy) \} &=& - \grad_\bfx \del(\bfx-\bfy).
	\label{e:PB-among-basic-var}
	\eeqs
Here $\om_{ij} = \pdr_i v_j - \pdr_j v_i$ is dual to vorticity, $w_i = \half \eps_{ijk} \omega_{jk}$ or $\omega_{lm} = \eps_{ilm} w_i$. 
$\{ v_i , v_j \}$ is akin to the PB of canonical momenta of a charged particle 
	\beq
	\left\{ p_i - ({e}/{c}) A_i(\bfx), p_j - ({e}/{c}) A_j(\bfx) \right\} = ({e}/{c}) F_{ij}(\bfx) 
	\eeq
where $F_{ij} = \eps_{ijk} B_k$. $\bfB$ is analogous to $\bfw$ and $F_{ij}$ to $\omega_{ij}$. 

The Jacobi identity $J = \{ \{ F, G \}, H \} + {\rm cyclic} = 0$ is formally expected if we regard (\ref{e:PB-among-basic-var}) as the semi-classical limit of Landau's commutators. However, it is not straightforward to verify in general [\onlinecite{morrison-aip}]. The Jacobi identity should also follow by interpreting (\ref{e:PB-fnals-of-A-rho-M}) as PBs among functions on the dual of a Lie algebra [\onlinecite{holm-kupershmidt}]. We have found a new direct proof of the Jacobi identity. It is first shown for linear functionals of $\rho$ and $\bfv$ using a remarkable integral identity that holds for arbitrary test fields $\bff, \bfg, \bfh$, any asymptotically constant $\rho$ and decaying $\bfw$:
	\begin{widetext}
	\beqs
	J &=& - \int \grad\left(\rho^{-2} \right) \cdot [(\bfw \cdot (\bff \times \bfg)) \bfh + (\bfw \cdot (\bfg \times \bfh)) \bff + (\bfw \cdot (\bfh \times \bff)) \bfg] \: d\bfr \cr
	&+& \int \frac{\bfw}{\rho^2} \cdot \left[ \left( \bff \times [\bfg, \bfh] + \bfg \times [\bfh, \bff] + \bfh \times [\bff, \bfg]  \right)  + \left\{ (\bfh \times \bfg) (\grad \cdot \bff) + (\bff \times \bfh) (\grad \cdot \bfg) + (\bfg \times \bff) (\grad \cdot \bfh)  \right\} \right] d\bfr = 0.
	\label{e:jacobi-expr-3-lin-fnals-of-v}
	\eeqs \normalsize
	\end{widetext}
The proof is extended to exponentials of linear functionals and then to wider classes of non-linear functionals via a functional Fourier transform. For details see Ref.~\onlinecite{arxiv-version}.

More generally, PBs among functionals of $\rho, \bfv$ and $\bfB$, for ideal compressible MHD [\onlinecite{morrison-and-greene}] are
	\beqs
	\label{e:PB-MHD}
	\{ F, G \} &=& \int \left[ \frac{\bfw}{\rho} \cdot F_\bfv \times G_\bfv - F_\bfv \cdot \grad G_\rho + G_\bfv \cdot \grad F_\rho \right] d\bfr \cr
	&& - \int (\bfB/\rho) \cdot \left[ \left( F_\bfv \cdot \grad \right) G_\bfB - \left( G_\bfv \cdot \grad \right) F_\bfB \right] \: d\bfr
	\cr
	&& + \int (B_i/\rho) \left[F_{v_j} \pdr_i G_{B_j} - G_{v_j} \pdr_i F_{B_j}  \right] d\bfr.
	\eeqs
In addition to the PBs among fluid variables (\ref{e:PB-among-basic-var}), it is remarkable that $\bfB$ commutes with $\rho$ and itself (in this it is unlike $\bfw$) while the PB of $\bfv$ with $\bfB$  is
	\beq
	\{ v_i(x), B_j(y) \} = \ov{\rho(x)} \eps_{ilk} \eps_{jmk} B_l(x) \pdr_{x^m} \del(x-y) .
	\label{e:pb-v-B}
	\eeq
For functionals of $\rho, \bfM = \rho \bfv$ and vector potential $\bfA$,
	\beqs
	&& \{ F , G \} = - \int \left[ 
	\bfM \cdot \left[ F_\bfM, G_\bfM \right]
	+ \rho \left( F_\bfM \cdot \grad G_\rho 
	- 	F \leftrightarrow G \right) \right] d\bfr
	\cr && - \int \bfA \cdot \left[ F_\bfM \grad \cdot G_\bfA - \grad \times \left( F_\bfM \times G_\bfA \right) - F \leftrightarrow G \right] d\bfr.
	\label{e:PB-fnals-of-A-rho-M}
	\eeqs
Thus components of $\bfA$ commute among themselves and with $\rho$, while the PB of $\bfA$ with velocity is
	\beq
	\{ v_i(x), A_j(y) \} = \frac{\left(\eps_{ijk}B_k(x) + A_i(x) \pdr_{y^j} \right)\del(x-y)}{\rho(x)} .
	\label{e:v-A-pb}
	\eeq
Remarkably, the standard PBs (\ref{e:pb-between-functionals-of-rho-v}, \ref{e:PB-fnals-of-A-rho-M}) of ideal Euler and MHD also imply the R-Euler and R-MHD equations if we pick the Hamiltonian as the conserved swirl energy
	\beq
	H = \int \left[ \frac{\rho \bfv^2}{2}+ U(\rho) + \frac{\la^2 \rho \bfw^2}{2} + \frac{\bfB^2}{2 \mu_0} \right] \: d\bfr
	\label{e:swirl-hamiltonian-MHD}
	\eeq
subject to the constitutive relation (\ref{e:constitutive-relation}) and the condition $U'(\rho) = h(\rho)$ (for adiabatic flow $U = p/(\gamma - 1)$). The $4$ terms in $H$ are kinetic (KE), potential (PE), enstrophic (EE) and magnetic (ME) energies ($\bfB$ = 0 in R-Euler).

For the continuity equation, we note that only KE contributes to $\{\rho, H \}$ since $\{ \rho, \rho \} = \{ \rho, \bfw \} = \{\rho, \bfB \} = 0$: \small
	\beq
	\rho_t = \{ \rho(\bfx), H \} 
	= \int_V \rho(\bfr) \bfv \cdot \grad_{\bfr} \del(\bfr-\bfx) \: d\bfr
	= - \grad \cdot (\rho \bfv).
	\eeq \normalsize
The surface term vanishes for $\bfx$ in the interior of $V$. For the momentum equation, we evaluate $\bfv_t = \{ \bfv, H \}$:
	\beqs
	&& \{ \bfv, KE \} = - (\bfv \cdot \grad) \bfv, \quad
	\{ \bfv, PE \} = - \grad U'(\rho),
	\cr
	&& \{ \bfv, EE \} = - \la^2 \bfT 
	\quad \text{and} \quad
	\{ \bfv, ME \} = (\bfj \times \bfB)/\rho.
	\eeqs
where $\mu_0 \bfj = \grad \times \bfB$. The surface terms vanish as before. The regularized equation for $\bfv$ then follows:
	\beq
	\bfv_t + \bfv \cdot \grad \bfv = - \grad U'(\rho) - \la^2 \bfw \times (\grad \times \bfw) + (\bfj \times \bfB)/\rho.
	\eeq
Formally, the R-Euler equations follow in a similar manner upon setting $\bfB, ME = 0$ above. Since both $\rho$ and $\bfB$ commute with $\bfA$, only KE and EE contribute to the evolution of the vector potential:
	\beq
	\pdr_t \bfA = \{\bfA, H \} = (\bfv_* \times \bfB) - \grad(\bfv_* \cdot \bfA). \label{e:pb-A-with-H}
	\eeq
We identify the electric field as $\bfE = - \pdr_t \bfA - \grad (\bfv_* \cdot \bfA)$. Thus in this `laboratory' gauge, the electrostatic potential is $\phi = \bfv_* \cdot \bfA$. This would be the electrostatic potential in the lab frame for the case where the electrostatic potential is zero in a `plasma' frame moving at $\bfv_*$ (See Eq.~24.39 of Ref.~\onlinecite{Fock}). In the lab frame, if $\bfv_* = 0$ at a point, then the electrostatic potential would be zero in this gauge at that point. This gauge is distinct from Coulomb gauge, indeed $\grad \cdot \bfA$ evolves according to
	\beq
	\pdr_t (\grad \cdot \bfA) = \grad \cdot (\bfv_* \times \bfB) - \grad^2 (\bfv_* \cdot \bfA).
	\eeq
Taking the curl of (\ref{e:pb-A-with-H}) we arrive at the regularized Faraday law governing evolution of $\bfB$
	\beq
	\pdr_t \bfB = \{ \bfB, H \}
	= \grad \times \left[ \bfv_* \times \bfB \right].
	\eeq
The Maxwell equation $\grad \cdot \bfB = 0$ is consistent with our PBs, for we verify that $\grad \cdot \bfB$ commutes with $H$. Since $\bfB$ commutes both with itself and with $\rho$, $PE$ and $ME$ cannot contribute to $\{ \grad \cdot \bfB, H \}$. On the other hand, one checks that KE and EE separately commute with $\grad \cdot \bfB$, so that it remains zero under hamiltonian evolution.

Conserved quantities and symmetry generators of compressible R-MHD [and R-Euler] satisfy a closed Poisson algebra. We briefly indicate a derivation of the conservation laws from the PB formalism. Using (\ref{e:constitutive-relation}), the PBs of linear (\ref{e:momentum-current-tensor}) and angular (\ref{e:ang-mom-current-conservation}) momenta with the {\it swirl} energy (\ref{e:swirl-hamiltonian-MHD}) can be expressed as fluxes of the corresponding currents across the boundary of the flow domain
	\beq
	\{ P_i , H \} = - \int_{\pdr V} \Pi_{ij} n_j dS, \quad
	\{L_i , H \} = - \int_{\pdr V} \Lambda_{il} n_l \: dS.
	\eeq
Thus $\{ P_i, H \} = \{ L_i, H \} = 0$ if these fluxes vanish at each point on the boundary (e.g. by specifying decaying BCs). The same BCs also imply that $\{ P_i, P_j \} = 0$ and that $\bfP$ and $\bfL$ transform as vectors under rotations: $\{ P_i , L_j \} = \eps_{ijk} P_k$ and $\{ L_i , L_j \} = \eps_{ijk} L_k$.

In R-Euler, the PB of the swirl hamiltonian with flow helicity $\{ H, {\cal K} \} = 0$ if we use $\bfw \cdot \hat n = 0$ and $\bfv \times \hat n = 0$ BCs.
Flow helicity also commutes with $\bfP$ and $\bfL$ with the same BCs. Indeed, $\cal K$ is a Casimir with these BCs. For, if $\bfv \times \hat n = 0$ then ${\cal K}_\bfv = 2 \bfw$ and for any $F$ functional,
	\beq
	\{ {\cal K}, F \} = -2 \int_{\pdr V} (\bfw \cdot \hat n) F_\rho dS = 0.
	\eeq
In R-MHD, $\{ {\cal K}_B, H \}$ is the flux of its current (\ref{e:mag-helicity-and-current-conservation}) in laboratory gauge ($\phi = \bfv_* \cdot \bfA$)
	\beq
	\{ {\cal K}_B, H \} = - 
	\int_{\pdr V} \left[(\bfA \cdot \bfB) \bfv_* \right] \cdot \hat n \: dS,
	\eeq
which vanishes if $\bfv_* \cdot \hat n = 0$ on $\pdr V$ (in other gauges we also need $\bfB \cdot \hat n = 0$). As with $\cal K$ in R-Euler, ${\cal K}_B$ is a Casimir in R-MHD. However, $\cal K$ is not conserved in R-MHD since the Lorentz force enters the momentum equation.

Finally, the Galilean boost generator ${\bf G} = \int \bfr \rho \: d\bfr$ is not conserved. Its PB with swirl energy is momentum
	\beq
	\{{\bf G}, H \} = \int \bfr \{ \rho, H \} d\bfr 
= - \int \bfr \grad \cdot (\rho \bfv) d\bfr	= \bfP
	\eeq
in both R-Euler and R-MHD. $\bf G$ transforms as a vector under rotations $\{ G_i, L_j \} = \eps_{ijk} G_k$ and there is a central term in $\{ G_i, P_j \} = M \del_{ij}$ where $M$ is the total mass of fluid. $\bf G$ of course commutes with $\cal K$ and ${\cal K}_B$ in R-Euler and R-MHD respectively.

An interesting application of our Hamiltonian and PB formulation is to the identification of other possible conservative regularizations that preserve Eulerian symmetries. These arise by choosing new constitutive relations. The twirl regularization $- \la^2 \bfT$ in R-Euler was picked as the least non-linear term of lowest spatial order preserving symmetries. With the constitutive relation (\ref{e:constitutive-relation}) it leads to a conserved swirl energy $E^*$, bounded enstrophy and a Hamiltonian formulation. Retaining the same PBs (Eq. \ref{e:PB-MHD}) as before, and choosing an unaltered form for the Hamiltonian,
	\beq
	H = \int \left[\half \rho \bfv^2 + U(\rho) + \half \la^2 \rho \bfw^2 \right] \; d\bfr,
	\label{e:hamiltonian-n-other-constitutive}
	\eeq
we will now allow for more general constitutive relations, e.g., $\la_n^2 \rho = c_n |\bfw|^{2n}$ where $c_n$ is a positive constant. The virtue of this type of constitutive law is that the $(n+1)^{\rm th}$ moment of $\bfw^2$ is bounded in the flow generated by this Hamiltonian. From Hamilton's equation for $\rho$ we see that the continuity equation is unaltered since $\rho$ commutes with itself and $\bfw$ (as long as $\la$ depends only on $\rho$ and $\bfw$, the continuity equation will remain the same). However, there is a new regularization term in the equation for $\bfv$. Indeed, the equation of motion $\bfv_t = \{ \bfv, H \}$ and continuity equation $\rho_t = \{ \rho, H \}$ become:
	\beqs
	&& \pdr_t \bfv + \bfw \times \bfv_{n*} = - \grad \sigma, 
	\quad \rho_t + \grad \cdot (\rho \bfv_{n*}) = 0
	\quad\text{where} \cr
	&& 	\bfv_{n*} = \bfv + \ov{\rho} \grad \times ((n+1) c_n|\bfw|^{2n} \bfw).
	\label{e:eom-new-constitutive-law}
	\eeqs
Thus the form of the governing equations is unchanged; only the swirl velocity $\bfv_*$ is modified. When $n = 0$, this reduces to the R-Euler equation with bounded first moment of $\bfw^2$ (enstrophy). For $n > 0$ we get more non-linear (of degree $2n+2$ in $\bfv$) regularization terms than the quadratic twirl term, though the equation {\it remains} $2^{\rm nd}$ order in space derivatives. Furthermore, $\bfP$ and $\bfL$ continue to be conserved as the new constitutive relation does not break translation or rotation symmetries (it only depends on the scalar $\bfw^2$). Flow helicity being a Casimir is still conserved, while parity, time reversal and Galilean boost invariance are also preserved. 

For R-MHD, the Hamiltonian (\ref{e:hamiltonian-n-other-constitutive}) is augmented by the magnetic energy term $\int (\bfB^2/2\mu_0) \: d\bfr$. We get the same R-MHD equations (\ref{e:R-MHD-Euler}, \ref{e:R-MHD-Faraday}) with $\bfv_*$ replaced by $\bfv_{n*}$. It is remarkable that the PB formalism enables us to obtain, with the help of a suitable constitutive relation, an arbitrarily strong a priori bound on vorticity.

\section{Steady R-Euler and R-MHD examples}
\label{s:examples}

\subsection{Columnar vortex solutions in R-Euler and R-MHD}
\label{s:modeling-vortex}

In this section we model a {\em steady} tornado [cylindrically symmetric rotating columnar vortex with axis along $z$] using the compressible R-Euler equations. The unregularized Euler equations do not involve derivatives of vorticity, and admit solutions where the vorticity can be discountinuous or even divergent (e.g. at the edge of the tornado). On the other hand, the R-Euler equations involve the first derivative of $\bfw$ and can be expected to smooth out large gradients in vorticity on a length scale of order $\la$ while ensuring bounded enstrophy.

In our rotating vortex model, $\rho, p$, $\bfv = v_\phi \hat \phi$ and $\bfw = w_z \hat z$ are all functions only of $r$, the distance from the axis of the columnar vortex. In the vortex core of radius $a$, we assume the fluid rotates at approximately constant angular velocity $\Omega$. Far from the core, $\bfw \to 0$. In a boundary layer of width $\ll a$, the $\bfw$ smoothly interpolates between its core and exterior values. The problem is to determine $\rho(r)$ given $w_z(r)$. As a consequence of the regularization term, we find that this decrease in vorticity is related to a corresponding increase in density (from a rare core to a denser periphery). By contrast, the unregularized Euler equations allow $\bfw$ to have unrestricted discontinuities across the layer while $\rho$ is continuous.

\begin{figure}[h]
\begin{center}
\fbox{\includegraphics[width = 4cm]{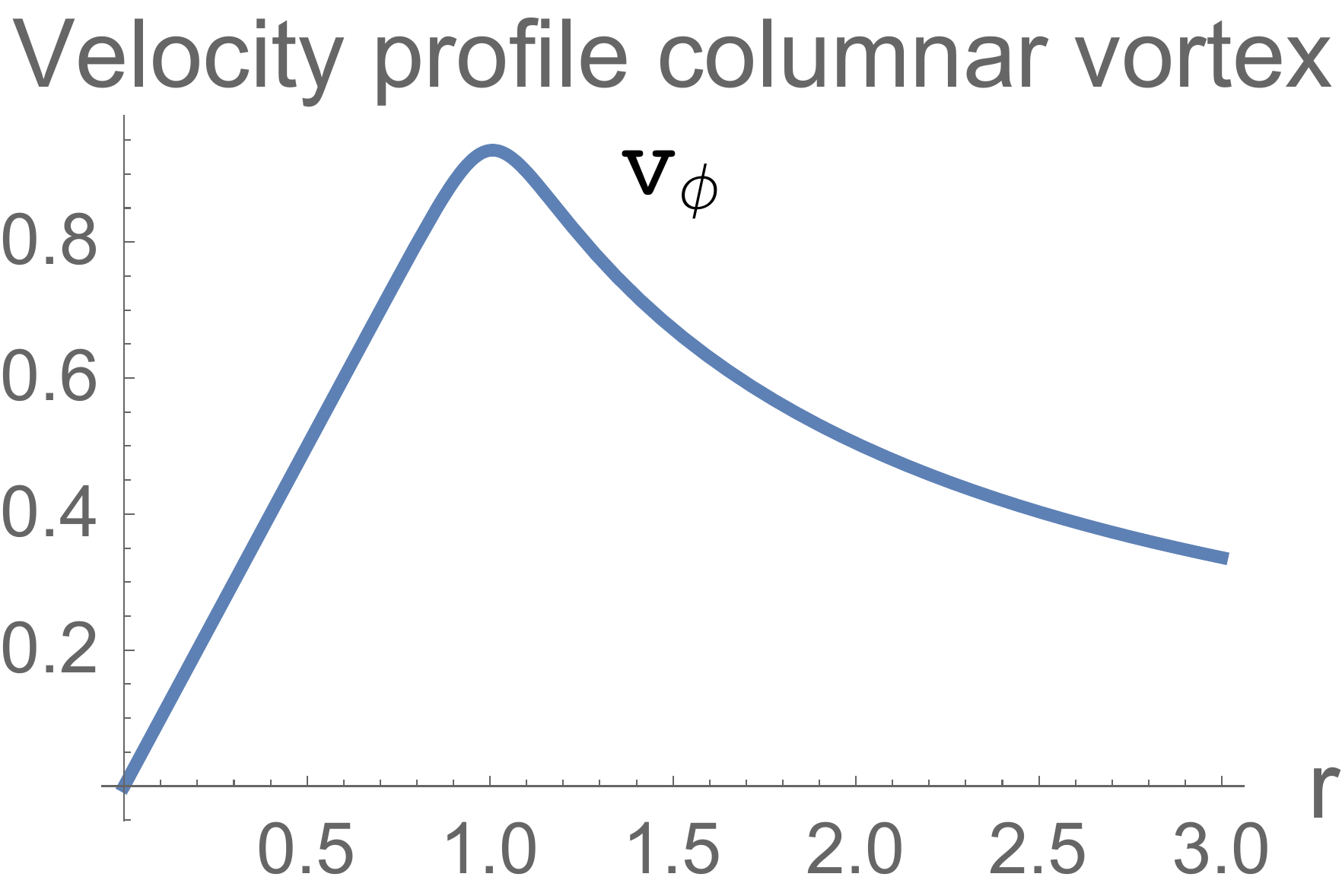}}
\fbox{\includegraphics[width = 4cm]{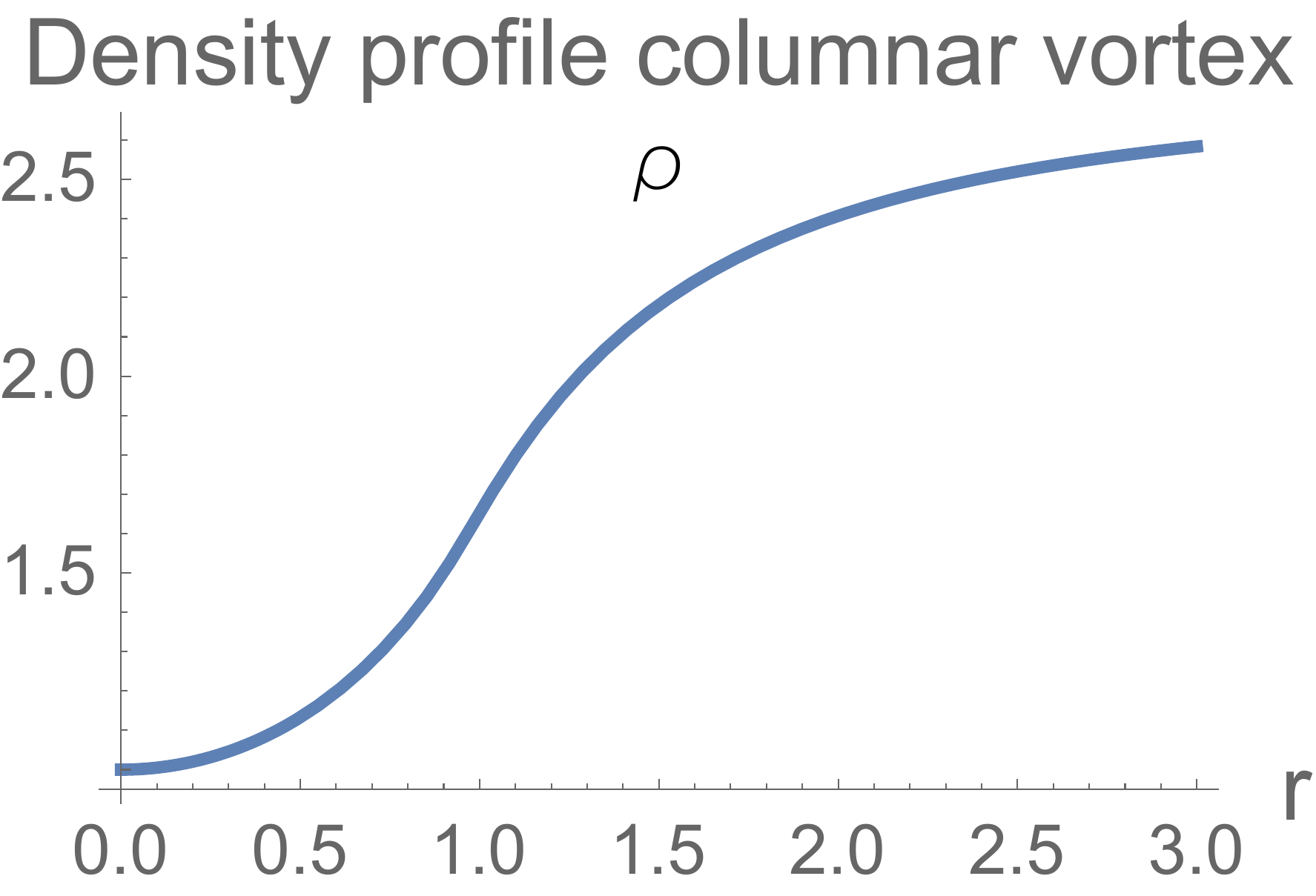}}
\caption{\small Velocity $v_\phi(r)$ and isothermal density $\rho(r)$ for rotating vortex of core radius $a=1$ and angular velocity $\Om=1$. Regularization relates drop in $w_z$ in a layer of thickness $\approx \eps = \lambda=0.1$ around $r = a$ to increase in $\rho$. The reference values are $p_0 = \rho_0 = 1$.}
\label{f:tornado-figs}
 \end{center}
\end{figure}

The steady continuity equation is identically satisfied. The steady state R-Euler equation (\ref{e:R-euler-v*-sigma}) has only a non-trivial radial component:
	\beq
	\frac{v_\phi^2}{r} = \dd{h}{r} + \frac{\la^2}{2} \dd{w_z^2}{r}.
	\label{e:steady-state-eqn-for-vortex}
	\eeq
We note that 
	\beq
	w_z = \ov{r} (r v_\phi)' \quad \text{and} \quad
	(\grad \times \bfw)_\phi = -w'(z).
	\eeq
As a simple model for a rotating vortex of core radius $a$, we consider the vorticity distribution (see Fig. \ref{f:tornado-figs})
	\beq
	w_z(r) = \frac{2 \Om }{\left[1+\tanh \left(a/\epsilon \right) \right]}\left[1 - \tanh \left(\frac{r-a}{\epsilon} \right) \right] .
	\label{e:vorticity-profile-tornado}
	\eeq
Over a transition layer of width $\approx 2 \eps \ll a$, the vorticity drops rapidly from $\approx 2 \Om$ to $\approx 0$. In the vortex core $r \ll a - \eps$, the flow corresponds to rigid body rotation at the constant angular velocity $\Omega \hat z$, apart from higher order corrections in $\epsilon$. Thus in the core, the vorticity is roughly twice the angular velocity and $\bfv = \Om \hat z \times \bfr \approx \Om r \hat \phi$. In the exterior region, for $r \gg a + \epsilon$ the vorticity tends to zero exponentially. The velocity $v_\phi = r^{-1} \int_0^r r w_z(r') dr'$ is obtained by integration. The velocity profile (Fig.\ref{f:tornado-figs}) rises nearly linearly with $r/a$ in the core [rigid body motion] and drops off as $\sim 1/r$ at large distances like a typical irrotational potential vortex. In the transition layer $a-\eps \lesssim r \lesssim a + \eps$ the radial derivative of the velocity varies rapidly.

The density can be obtained by integrating the steady R-Euler equation. We do this below in the simple case of isothermal flow ($p = (p_0/\rho_0) \rho$) where $h = p_0/\rho_0 \ln(\rho/\rho_0)$. The adiabatic case ($p/p_0 = (\rho/\rho_0)^\gamma$) is similar, but (\ref{e:steady-state-eqn-for-vortex}) is a non-linear first order ODE for density which can be integrated numerically.

The steady equation (\ref{e:steady-state-eqn-for-vortex}) in the isothermal case is
	\beq
	\frac{p_0}{\rho_0} \rho'(r) - \frac{v_\phi^2}{r} \rho(r) = - \frac{\la_0^2 \rho_0}{2} (w_z^2)'.
	\label{e:isothermal-steady-state-vortex-eqn}
	\eeq
It is convenient to take the reference values $\rho_0, \la_0, p_0$ to be at $r = 0$. The solution for $\rho(r)$ is
	\beqs
	\rho 
	&=& \frac{\rho_0 q(0)}{q(r)} \left[ 1 - \frac{\Om^2 \la_0^2 \rho_0}{2 p_0} \int_0^r \frac{q(s)}{q(0)} \left(\frac{w_z^2}{\Om^2} \right)' \, ds \right] \cr
	\text{where} &&
	\frac{q(r)}{q(0)} = \exp \left[- \frac{\rho_0}{p_0} \int_0^r \frac{v_\phi^2}{s} \, ds \right].
	\eeqs
$q(r)$ is a positive monotonically (exponentially) decreasing function of $r$ and we can take $q(0)=1$ without loss of generality. The integrations are done numerically and the resulting density is plotted in Fig \ref{f:tornado-figs}. $\rho$ is monotonically increasing from $\rho(0)$ to an asymptotic value $\rho(\infty)$ (material has been `ejected' from the core). The above formula shows that one effect of the regularization is to decrease the density relative to its Eulerian value (especially outside the core). To get more insight into the role of the regularization we solve the steady equation approximately in the core, transition and exterior regions separately.


{\flushleft \bf Vortex Core $0 < r \lesssim a_- = a-\eps$}: In this region $w_z(r) \approx w_z(0) = 2\Omega$. The corresponding velocity $v_\phi(r) = r w_z(0)/2=r\Omega$ grows linearly as for a rigidly rotating fluid. The density grows exponentially inside the vortex core, for $r \lesssim a_-$:
	\beq
	\rho(r) \approx \rho(a_-) \exp \left[\rho_0 \Om^2 (r^2 - a_-^2)/2 p_0 \right]. 
	\eeq


{\flushleft \bf Outside the vortex $r \gtrsim a_+ = a+\eps$:} Here $w_z(r) \approx 0$ so the velocity decays as $v_{\phi}(r) = a_+ v_\phi(a_+)/r$. Again, ignoring the regularization term, the steady state density is determined by (\ref{e:isothermal-steady-state-vortex-eqn})
	\beq
	\frac{\rho'(r)}{\rho(r)} = \frac{\rho_0 a_+^2 v_\phi(a_+)^2}{p_0} \ov{r^3}.
	\eeq
$\rho(r)$ monotonically increases from its value at the outer edge $\rho(a_+)$ to an asymptotic value $\rho(\infty)$ 
	\beq
	\rho(r)	= \rho(a_+) \exp \left[ \frac{\rho_0 v_\phi(a_+)^2 \left(r^2 - a_+^2 \right)}{2 p_0 r^2} \right].
	\eeq
Even in this approximation, $\rho$ in the exterior depends on the regularization via $v_\phi(a_+)$.


{\flushleft \bf Transition layer $a_- \lesssim r \lesssim a_+$:} Here $w_z(r)$ (\ref{e:vorticity-profile-tornado}) rapidly falls from $w_z(0)$ to $0$. 
$\rho$ is determined by 
	\beq
	\frac{\rho v_\phi^2}{r} = \frac{p_0}{\rho_0} \rho'(r) + \frac{\la^2 \rho}{2} \dd{w_z^2}{r}.
	\label{e:steady-R-Euler-for-rot-vortex}
	\eeq
To find the density we integrate this equation from $a_-$ to $r < a_+$ using the relation $\la^2 \rho = $ constant:
	\beq
	\int_{a_-}^r \frac{\rho v_\phi^2}{r'} dr' = \frac{p_0}{\rho_0} \left[ \rho(r) - \rho(a_-) \right] + \frac{\la^2 \rho}{2} \left[ w_z^2(r) - w_z^2(a_-) \right].	
	\eeq
Since the layer is thin ($\eps \ll a$) and $\rho$, $v_\phi$ are continuous across the layer, we may ignore the LHS. Thus the rapid decrease in $w_z$ must be compensated by a corresponding increase in $\rho$ across the layer
	\beq
	(2p_0/\rho_0) \left[ \rho(r) - \rho(a_-) \right] \approx {\la^2 \rho} \left[w_z^2(a_-) - w_z^2(r) \right].
	\label{e:density-vorticity-balance-across-layer}	\eeq
The increase in $\rho$ is not as rapid as the fall in $w_z$ since the latter is multiplied $\la^2$. For our vorticity profile (\ref{e:vorticity-profile-tornado}), taking $w_z(a_-) \approx w_z(0) = 2 \Om$, we get $\rho(r)$ in the transition layer
	\beq
	 \rho(r) \approx \rho(a_-) + \frac{2 (\Om \la_0)^2 \rho_0^2}{p_0} \left[1 - \frac{(1 - \tanh((r-a)/\eps))^2}{(1 + \tanh(a/\eps))^2} \right].	\eeq
In particular, $\rho(a_+)$ exceeds $\rho(a_-)$ by an amount determined by the regularization
	\beqs
	\rho(a_+) &\approx& \rho(a_-) + \frac{2 (\Om \la_0)^2 \rho_0^2}{p_0} \left[1 - \frac{[1 - \tanh(1)]^2}{(1 + \tanh(a/\eps))^2} \right]
	\cr
	&\approx& \rho(a_-) + 2 M^2  \rho_0 \quad \text{for} \quad \eps \ll a.
	\eeqs
We see that for $\eps \ll a$ (vortex edge thin compared to core size), the twirl force causes an increase in density across the boundary layer by an amount controlled by the `twirl Mach number' $M = \la_0 \Om/c_s$ where $c_s = \sqrt{p_0/\rho_0}$ is the isothermal sound speed.

The steady R-Euler equation (\ref{e:steady-R-Euler-for-rot-vortex}) for the vortex is similar to Schr\"odinger's stationary equation for a non-relativistic quantum particle in a 1d delta potential: $E \psi(x) = - g \del(x) \psi(x) - ({\hbar^2}/{2m}) \psi''(x)$. $E \psi$ is like ${\rho v_\phi^2}/{r}$ on the LHS of (\ref{e:steady-R-Euler-for-rot-vortex}). The potential $- g \del(x) \psi(x)$ and kinetic $-(\hbar^2/2m) \psi''(x)$ terms mimic the pressure $(p_0/\rho_0) \rho'$ and twirl $({\la^2 \rho}/{2}) (w_z^2)'$ terms respectively. The kinetic and twirl terms are both singular perturbations. The free particle regions $x < 0$ and $x > 0$ are like the interior and exterior of the vortex. The bound-state wave function is $\psi(x) = A \exp(- \kappa |x|)$ with $\kappa = \sqrt{-2mE}/\hbar$, so $\psi'$ has a jump discontinuity at $x=0$. The boundary layer is like the point $x=0$ where the delta potential is supported. Just as we integrated R-Euler across the transition layer, we integrate Schr\"odinger in a neighbourhood of $x=0$ to get $\psi'(\eps) - \psi'(-\eps) = - (2mg/\hbar^2) \psi(0)$. The discontinuity in $\psi'$ is determined by $\psi(0)$, just as the increase in $\rho$ across the layer is fixed by the corresponding drop in $w_z$ (\ref{e:density-vorticity-balance-across-layer}). Finally, $\la > 0$ regularizes Euler flow just as $\hbar > 0$  regularizes the classical theory, ensuring $E_{\rm gs} = - {m g^2}/{2 \hbar^2}$ is bounded below.


{\flushleft \bf A columnar vortex in conjunction with an MHD pinch:} A similar analysis in R-MHD involves specifying in addition to the above, $j_z(r)$ and $B_\phi(r)$ associated with it. Thus the radial momentum equation in R-MHD under isothermal conditions becomes
	\beq
	\frac{p_0}{\rho_0} \rho' - \frac{v_\phi^2}{r} \rho = - \half \la_0^2 \rho_0 (w_z^2)' - \frac{B_\phi}{\mu_0 r} (r B_\phi)'
	\label{e:mhd-pinch-vortex-ODE-for-rho}
	\eeq 
where $\mu_0 j_z = r^{-1} (r B_\phi)'$. The R-Faraday equation $\grad \times (\bfv_* \times \bfB) = 0$ is identically satisfied since both $\bfv_* = (v_\phi - \la^2 w_z') \hat \phi$ and $\bfB$ are parallel. Thus the electric field is zero. In (\ref{e:mhd-pinch-vortex-ODE-for-rho}) the inhomogeneous term on the RHS is modified by the Lorentz force. The latter is always radially inwards (`pinching') whereas the twirl term is outwards for radially decreasing vorticity and furthermore could be small for $\la_0 \ll a$. Thus the radial density variation in this magnetized columnar pinch could differ from R-Euler. For any given current and vorticity profiles (\ref{e:mhd-pinch-vortex-ODE-for-rho}) can be integrated to find $\rho(r)$.

Another case of interest in R-MHD is a magnetized columnar vortex with an axial skin current. Thus we assume $j_z(r)$ is localized between $a - c/\om_{pe}$ and $a + c/\om_{pe}$ where $c/\om_{pe}$ is the electron collisionless skin depth and $\la \approx c/\om_{pe}$. In this case, in the interior $r < a_-$ we have the previous (tornado) interior solution with $B_\phi = 0$. In the exterior solution, $B_\phi(r) \approx \mu_0 I/2\pi r$ for $r \geq a_+$. The effect of the Lorentz force in the skin is seen from (\ref{e:mhd-pinch-vortex-ODE-for-rho}) to be opposite to that of the twirl term. The exclusion of the magnetic field within the vortex is reminiscent of the Meissner effect in superconductivity. Axial fields (screw pinch) and flows with the same symmetries may be readily incorporated in the framework presented.

\subsection{Isothermal plane vortex sheet}

Consider a steady plane vortex sheet of thickness $\tht$ lying in the $x$-$z$ plane. Assume the velocity points along $x$, $\bfv = (u(y),0,0)$ and approaches {\em different} asymptotic values $u_\pm$ as $y \to \pm \infty$. $\rho$ is also assumed to vary only with height $y$. The steady continuity equation is identically satisfied. For our velocity field the advection term $\bfv \cdot \grad \bfv \equiv 0$. Denoting derivatives by subscripts,
	\beqs
	&& \bfw = - u_y \hat z, \quad
	\bfw \times \bfv = - u u_y \hat y, \quad
	\grad \times \bfw = - u_{yy} \hat x
	\cr
	&& \text{and} \quad
	\bfT = \bfw \times (\grad \times \bfw) = u_y u_{yy} \hat y.
	\eeqs
Only the $\hat y$ component of the R-Euler equation survives:
	\beq
	\la^2 u_y u_{yy} = - \pdr_y{h(\rho(y))}.
	\label{e:R-euler-vortex-sheet}
	\eeq
For isothermal flow, specific enthalpy is $h = (p_0/\rho_0) \log(\rho/\rho_0)$. Using (\ref{e:constitutive-relation}), (\ref{e:R-euler-vortex-sheet}) becomes
	\beq
	\pdr_y \left(\half \la^2 \rho u_y^2 + \frac{p_0 \rho}{\rho_0} \right) = 0.
	\label{e:R-euler-isotherm-vortex-sheet-bernoulli}
	\eeq
The steady state is not unique. (\ref{e:R-euler-isotherm-vortex-sheet-bernoulli}) can be used to find $\rho(y)$ for any given vorticity profile. (\ref{e:R-euler-isotherm-vortex-sheet-bernoulli}) can be loosely regarded as a regularized version of Bernoulli's equation: the sum of enstrophic and compressional energy densities is independent of height. The kinetic contribution is absent for a longitudinal velocity field varying only with height (the advection term is identically zero). This Bernoulli-like equation is very different from the usual one, which involves kinetic and compressional energies. In that case, the pressure is lower where the velocity is higher. In the present case, we find that the density, and hence the pressure, is lower where the vorticity is higher. This is fundamentally a consequence of the regularizing ``twirl acceleration''.

To model a vortex sheet of thickness $\tht$ we take the vorticity profile in $y$ to be given by
	\beq
	u_{y} = \Delta u \: \left(\frac{\theta}{\pi} \right) \left[\frac{1}{\theta^{2}+y^{2}} \right] \quad \text{where} \quad \bfw = - u_y(y) \: \hat z
	\label{e:vortex-sheet-vel-profile}
	\eeq
Here $\Delta u = u_{+} - u_{-}$ and $w_{0} = -{\Delta u}/{\pi \theta}$ is the $z$-component of vorticity on the sheet. We obtain the first integral,
	\beq
	\half \la_{0}^2 \rho_{0} u_y^2 + \frac{p_{0} \rho}{\rho_0} = K.
	\eeq
The suffix in this instance refers to quantities on the sheet ($y=0$). The Bernoulli constant
	\beq
	K = p_0 + \half \rho_0 (\Delta u)^2 \, \left( \frac{\la_0}{\pi \tht} \right)^2.
	\eeq
We obtain the velocity profile by integrating (\ref{e:vortex-sheet-vel-profile}):
	\beq
	u(y) = u_{-} + (\Delta u) \left[\half + \frac{1}{\pi}\arctan\left(\frac{y}{\theta} \right) \right].
	\eeq
Assuming $u_+ > u_-$, the velocity monotonically increases from $u_-$ to $u_+$ with increasing height $y$. Moreover, the velocity on the sheet $u(0) = (u_- + u_+)/2$ is the average of its asymptotic values. The density profile follows from the first integral:
	\beq
	 \frac{\rho}{\rho_0} = 1 + \left(\frac{\la_0 }{\pi \theta} \right)^2 \left[\frac{\rho_0 (\Delta u)^{2}}{2p_0} \right] \left[1- \left(\frac{\theta^{2}}{\theta^{2}+y^{2}} \right)^{2} \right].
	\eeq
In particular, the asymptotic densities are
	\beq
	\frac{\rho_{\pm \infty}}{\rho_0} = 1 + \half \left( \frac{\la_0}{\pi \theta} \right)^2 \left[\frac{\rho_0 (\Delta u)^2}{p_0} \right].
	\eeq
Thus, the density is decreased at the sheet relative to the values at $\pm \infty$. If the sheet thickness $\theta \gg \la_0/\pi$, the decrease is not significant. If the thickness is comparable to the regularizing length $\la_0$, the density decrease at the sheet can be considerable, depending upon the `relative flow Mach number' defined as, $(\Delta M)^{2} = (\rho_0/p_0)(\Delta u)^2$. Unlike velocity, the density increases from the sheet to the same asymptotic values on either side of the sheet ($y = \pm \infty$), reflecting the symmetry of the assumed vorticity profile. This is similar to the rotating vortex/tornado model (\S\ref{s:modeling-vortex}) where an increase in density outwards from the core of the vortex is balanced by a corresponding decrease in vorticity. 

\section{Discussion}
\label{s:discussion}

The motivation and issues arising in regularizing conservative, continuum systems like Eulerian ideal fluid mechanics and ideal MHD were explained with some examples in Ref.~\onlinecite{thyagaraja}. Here we take up some points relevant to the present work.
We note that kinetic approaches such as the Chapman-Enskog method based on, for example the Fokker-Planck equation of plasma theory, typically lead in higher orders in the mean-free-path asymptotic expansion to both ``entropy conserving reactive''  and dissipative terms in the stress tensor and the heat-flux vector [\onlinecite{Braginskii}, \onlinecite{LifPit}]. It is possible that terms like the ``twirl-acceleration'' [introduced here essentially as a formal conservative regularizing effect] could arise in higher order asymptotics [like Burnett expansion] of kinetic equations.
Our work is based on the principle that singularities such as unbounded enstrophies in ideal MHD and neutral fluids and/or finite time failure of the models [\onlinecite{henneberg-cowley-wilson}] should be removed, if possible, by suitable {\it local} regularizing terms in the governing equations, in the spirit of Landau's mean field theory as discussed in the introduction. 

In the present paper, we have obtained compressible R-Euler and R-MHD equations which have a positive-definite energy density. It is worth noting that unlike driven dissipative systems like NS and visco-resistive MHD, in our conservative models the number of effective degrees of freedom and recurrence properties are determined by {\it initial data} [\onlinecite{lashmore-mccarthy-thyagaraja}].

We have shown that the swirl energy (\ref{e:swirl-energy-R-Euler}, \ref{e:swirl-hamiltonian-MHD}) is a constant of the motion and thus implies an a priori bound for enstrophy and energy. The system motion takes place in the function space of $\rho(\bfx),\bfv(\bfx)$ which is ``foliated'' by the closed, nested hyper-surfaces formed by the constant energy. The models are shown to be time reversible and to satisfy the symmetries of the Euler equations and have corresponding conservation laws. We have deduced Kelvin-Helmholtz-Alfv\'en-type ``frozen-in'' theorems associated with the swirl velocity $\bfv_*$ (\S\ref{s:swirl-vel-and-intergral-invariants}). We have demonstrated the remarkable fact that the R-Euler and R-MHD models are Hamiltonian with respect to the same Landau-Morrison-Greene [\onlinecite{landau}, \onlinecite{morrison-and-greene}] Poisson brackets previously derived for the unregularized models.

A significant application of the PB formalism is a generalization of the simple constitutive relation $\la^2 \rho = $ constant for compressible flows to a wider class of conservatively regularized models with a priori bounds on higher moments of vorticity.

It is useful to note that a possible approach to the statistical mechanics of R-Euler/MHD systems is through the Hopf distribution functional [\onlinecite{stan}]. Although originally conceived as a method of investigating the statistical theory of NS turbulence, the Hopf functional can certainly be of value in the regularized conservative models. Thus our PBs allow us to formulate Hopf's equation (analogue of the Liouville equation) $F_t + \{ F, H \} = 0$ for the functional $F[\rho, \bfv,t]$. Moreover, the Hamiltonian structure of the flow on the energy hyper-surface leads to micro-canonical statistical mechanics, and more generally to a canonical distribution.

A statistical mechanics of entangled 3D regularized vortex/magnetic flux tubes with bounded enstrophy and energy in dissipationless compressible motion would be a significant extension of the 2D theory of line vortices and filaments [\onlinecite{KRSreenivasan-onsager}, \onlinecite{edwards-taylor}].

As noted, NS can be regularized by adding a `hyper-viscosity' that depends on velocity gradients [\onlinecite{ladyzhenskaya}]. We conjecture that it may be possible to demonstrate the existence of unique classical solutions of NS and visco-resistive MHD regularized with our twirl term. This is based on the locally non-linear conservative nature of the twirl term which balances the vortex stretching term in analogy with hyperviscosity. However, this problem is outside the scope of this work.

It is interesting to mention that a 1D analogue of the twirl-regularized visco-resistive MHD model is the KdV-Burgers equation investigated by Grad and Hu (in Refs.~\onlinecite{Grad-Hu} and \onlinecite{Hu-KdV-Burgers}) in the context of weak plasma shocks propagating perpendicular to a magnetic field (electron inertia effects on weak non-linear plasma waves). 

The ideas due to Koopman and von Neumann [\onlinecite{RieszNagy}] in ergodic theory are also directly relevant provided a suitable measure can be developed for the constant energy surface on which the system motion takes place. The possibility of mapping the nonlinear R-Euler evolution to a 1-parameter group of unitary transformations in a function space of effectively finite number of degrees of freedom could have many practical applications. 

In numerical simulations of conservative systems it is crucial to monitor the quality of the calculation by careful evaluation of the conserved quantities. Thus having a conserved positive definite Hamiltonian and an a priori bound on enstrophy are powerful tools to control the micro-scale behavior of the dynamics and evaluate, on all scales, the relative sizes of energy and enstrophy. Unlike in dissipative systems like NS which are associated with semi-groups, our models involve 1-parameter groups of transformations generated by the Hamiltonian through the PBs. This has important implications for the implementation of numerical schemes for time evolution. 
Our examples show that the regularization can effectively remove effects arising from discontinuities in velocity derivatives. The vortex sheet suggests that the density near the sheet is always lowered relative to asymptotic, far-field values, just as the density in the core of our rotating tornado is lower than outside. However, in the corresponding R-MHD case we find that the magnetic field tends to increase the core density due to the pinch effect. The Kelvin-Helmholtz and current-driven instabilities of regularized vortex/current sheets/filaments and rotating vortices is of considerable interest. The a priori bound on enstrophy and energy demands a purely conservative non-linear saturation of any linearly growing mode. The behavior of such nonlinear dynamics could provide insight into the statistics and kinematics of turbulent motions in the inertial range.

Incidentally, all continuous potential flows of standard Euler theory in which ${\bf w} \equiv 0$ are also solutions of R-Euler. In otherwise irrotational flow, it is only when vortical singularities are encountered, that our theory differs by regularizing the solutions. However, it must be stressed that the twirl force cannot resolve all singularities of inviscid gas dynamics and ideal MHD. A simple example is provided by the plane normal shock. Taking $\rho(x),u(x)$ and $p(x)$ as the basic variables in 1D, clearly at the shock front, these quantities change rapidly. However, no vorticity is associated with the flow and the twirl force is absent. It is well-known that collisional shock fronts involve entropy rises. Thus, to regularize them one could add viscosity. On the other hand, to deal with collision-less shocks one could extend the swirl Hamiltonian to include $(\grad \rho)^2$-type terms. 

To sum up, our 3D regularized systems, while not conserving energy and enstrophy separately (unlike in 2D) do allow for both of them to be bounded a priori through non-linear dispersive interactions. This is achieved using a Hamiltonian structure based on the conserved positive definite swirl energy (c.f. Eq. \ref{e:swirl-hamiltonian-MHD}).

\vspace{-.7cm}
\begin{acknowledgments} 
\vspace{-.3cm}

We thank M Birkinshaw, R Nityananda, S G Rajeev, J Samuel, A Sen and A Young for stimulating discussions and a referee for suggesting improvements. Support of CMI for AT is acknowledged. This work was supported in part by the Infosys Foundation and a Ramanujan grant.

\end{acknowledgments}

\appendix


\end{document}